\documentclass{elsart1p}
\usepackage{graphicx}
\usepackage{lineno}
\usepackage{amssymb}
\begin{document}
\begin{frontmatter}
%
%
%
%
%
\title{Open Charm Analysis at Central Rapidity in ALICE using the  first year of  pp data at $\sqrt{s}$= 7 TeV }
%
%

\author{Renu Bala, for the ALICE Collaboration} 
\ead{bala@to.infn.it}
\address{INFN Sezione di Torino}

\begin{abstract}
ALICE is the dedicated heavy-ion experiment at the LHC. Its main
physics goal is to study the properties of the strongly-interacting
matter in the conditions of high energy density ($>$10 GeV/$\rm fm^{3}$)
and high temperature ($>$ 0.3 GeV) expected to be reached in central
Pb--Pb collisions. Charm and beauty quarks are a powerful tool to
investigate this high density and strongly interacting state of matter as they are produced in initial hard scatterings, and due to their long life
time, they probe all the stages of the system evolution. 
The detector design was optimized for heavy ions but
is also well suited for pp studies. ALICE recorded pp data
at $\sqrt{s}$= 7 TeV since march 2010 and the first run with heavy ion
collisions took place in November 2010.  The measurement of  charm production cross section in pp collisions provides
interesting insight into QCD processes and is important as a
reference for heavy ion studies. The measurement of the D-meson yield
in pp collisions can be used to extract the charm cross section. In
this contribution, the ongoing study of reconstruction of D-mesons
through  hadronic decay channels and the first preliminary results obtained with $\sqrt{s}$= 7 TeV pp data will be presented. 
\end{abstract}

\end{frontmatter}

\section{Physics Motivation}

Heavy quarks are unique probes to
study the Quark-Gluon Plasma produced in heavy ion collisions at the LHC.  Due to their large masses, they are produced
predominantly  in hard scatterings,  during the
initial phase of the collision. Therefore, they can probe the
properties of the nuclear matter created during the entire space time
evolution. One of the key methods used to infer the parameters of the matter is the measurement of energy loss of the partons traversing
it. Heavy quarks are expected to lose less energy in nuclear matter
than light quarks and gluons because of the mass dependent suppression
of the gluon radiation at small angles (dead-cone effect)
\cite{dead-cone}.
Heavy-quark production measurements in pp collisions at the LHC, besides providing a necessary
reference for the study of medium effects in Pb--Pb collisions, are interesting per se, as a
test of perturbative QCD (pQCD) in a new domain, up to 7 times above the present energy frontier
at the Tevatron. The charm production cross section measured in
$p\bar{p}$ collisions at $\sqrt{s}$=1.96 TeV at the Tevatron \cite{CDF}
is on the upper limit of the pQCD
caculations, as 
observed also in pp collisions at RHIC at much lower energy of $\sqrt{s}$=
0.2 TeV \cite{STAR}.

\section{ALICE Detector}
The ALICE detector \cite{JINST}  consists of two parts:  a central
barrel at central rapidity  and a muon spectrometer at forward
rapidity. For the  present analysis,  we have used the information from a
subset of the central barrel detector, namely the Inner Tracking System
(ITS), the Time Projection Chamber (TPC), the Time Of Flight
detector (TOF), T0 for time zero measurement  and V0 scintillator for triggering. The two tracking
detectors, the ITS  and the TPC, allow the reconstruction of 
charged particle tracks in the pseudorapidity range -0.9 $< \eta <
$0.9 with a  momentum resolution better than 2$\%$ for $p_{t} < $20 GeV/c and provide particle identification via dE/dx measurement.
The ITS, in particular,  is a key detector for open heavy flavour studies because it
allows to measure the track impact parameter (i.e. the distance of
closest approach of the track to the primary vertex) with a resolution
better than 75 $\mu$m for $p_{t} > $ 1 GeV/c  thus providing
the capability to detect the secondary vertices originating from
heavy-flavour decays. The TOF detector provides  particle identification by
time of flight measurement.
The results that we present are  obtained from a sample of $\rm 10^{8}$ min. bias
events triggered with the V0 scintillator and SPD detector (two
innermost layers of ITS detector), which corresponds to  20 $\%$ (1.4$\rm nb^{-1}$)  of the total 2010
statistics.
\section{Measurement of Charm Production Cross Section} 
Here we will discuss the strategy for cross section measurement for
two hadronic decay channels, i.e $\rm D^{0} \rightarrow K^{-} \pi^{+}$ and
$\rm D^{+} \rightarrow K^{-} \pi^{+} \pi^{+}$.
The analysis strategy is based on an invariant mass analysis of fully
reconstructed decay topologies originating from displaced
vertices. The cross section is calculated from the raw signal yield
extracted with invariant mass analysis using the following formula: 
\begin{equation}
\label{eq:cross}
\left. \frac{d\sigma}{\rm dp_{t}}\right|_{|y|<0.5}= \frac{1}{2} \cdot
\frac{1}{\rm \Delta y(p_{t})}
\cdot \frac{1}{BR} \cdot \frac{1}{\epsilon_{c}}\cdot f_{c}(p_{t})
\cdot \frac{\rm N^{D}_{Raw}(\rm p_{t})|_{|y|<\rm \Delta y(p_{t})}}{\rm N^{tot}_{MinBias}} \cdot \sigma^{tot}_{MinBias}
\end{equation}

The different terms in the above equation are described in the
following steps.
\begin{itemize}
\item[$\bullet$] Raw Signal Extraction ($\rm N^{D}_{Raw}(p_{t})$). Due to large
combinatorial background, the topological cuts have been tuned and
applied in order to maximize the statistical significance. In the case
of $\rm D^{0}$ mesons, the two main cut variables are the product of the impact parameters of the two
tracks ($d^{K}_{0} \times d^{\pi}_{0}$) and the cosine of the pointing
angle ($\theta_{pointing}$). 
For the $\rm D^{+}$ meson, 
the two main selection variables are the distance
between the reconstructed primary and secondary vertices and the
cosine of the pointing angle. Particle identification, provided by
TOF and TPC, helps to further reduce the background at low $p_{t}$.
\item[$\bullet$] Correction efficiency ($\epsilon_{c}$). The measured yield is
then  corrected for detector acceptance and cut selection efficiency
extracted from MC simulation with detailed description of detector
response and experimental conditions.
\item[$\bullet$] Correction for feed down from B-mesons ($f_{c}$). A relevant fraction of D's
comes from B-meson decays and since the tracks coming from secondary D are
well displaced from the primary vertex, due to the  relatively
long life time of B-mesons (c$\tau \approx$ 460-490 $\mu$m), the
selection further enhances their contribution to the raw yield (upto $\approx$15
$\%$). This contribution must be
subtracted. We have done this estimation using the beauty production
cross section predicted by FONLL calculation \cite{cacciari} and the detector simulation.  
We will check this estimation with the data driven method developed by
CDF Collaboration \cite{CDF} with the full 2010
statistics. 
\item[$\bullet$] Cross section normalization:  The corrected yield is then
divided by the branching ratio, by the acceptance in rapidity ($\rm \Delta
y(p_{t})$)  of
each $p_{t}$ interval and by a factor 2 (as both particle and
antiparticle are measured). Then,  the resulting value  is divided by the
integrated luminosity to obtain the $p_{t}$ differential cross section. 
\end{itemize}
 Each term of the above formula has some systematic
 uncertainty. The main sources of systematic errors are: raw yield
 extraction (10 $\%$) which was determined by repeating the fit, in
 each $ p_{t}$
 interval, in a different mass range and also with a different
 function to describe the background and feed down from B ($\approx$
 15 $\%$). The systematic error for  correction factors appplied
 are also taken into account. Considering all the sources of systematic,
 we have 20-25$\%$ of total systematic uncertainties. 
\begin{figure}[ht]
\includegraphics[width=16pc,height=14pc]{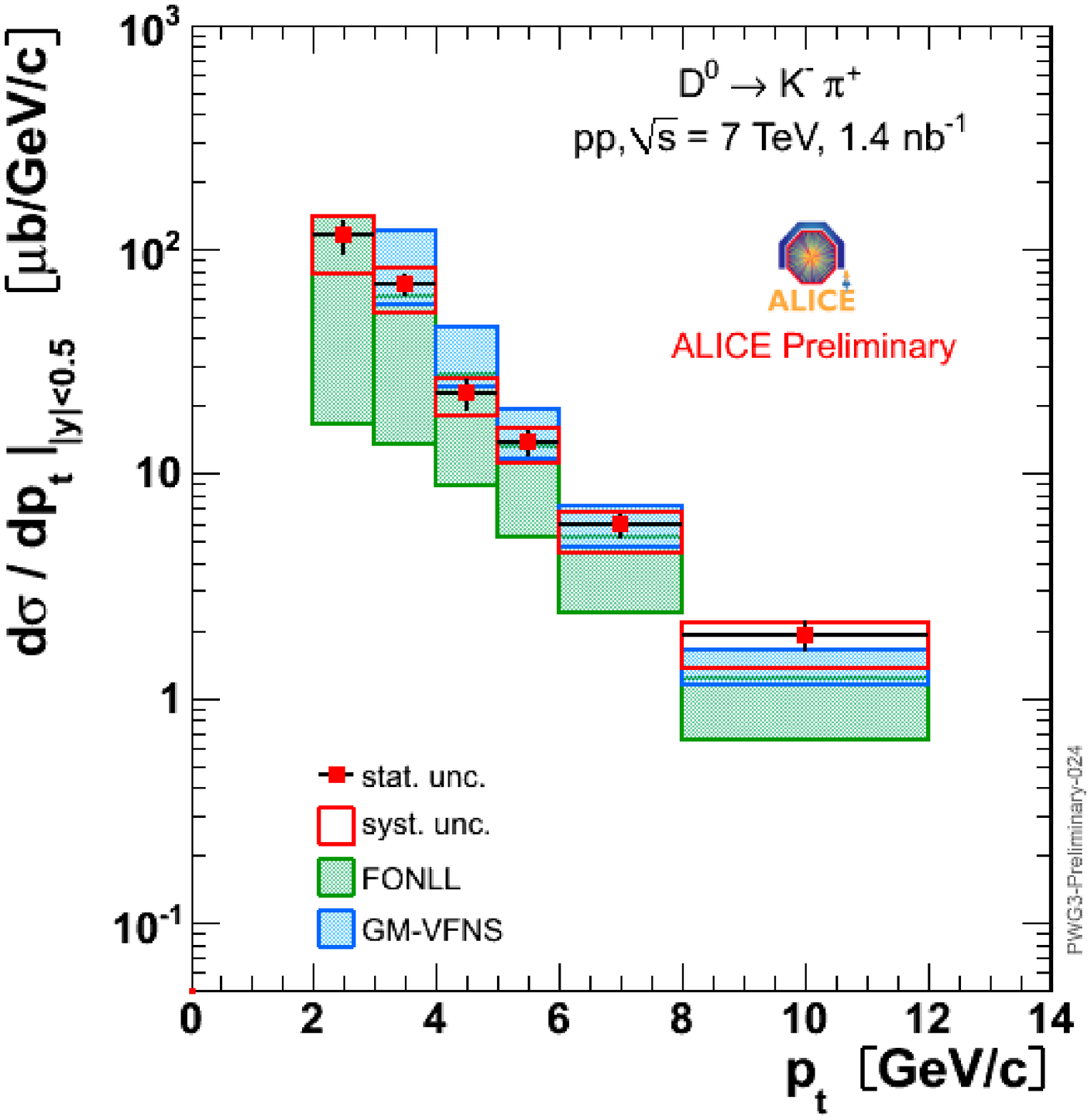}
\includegraphics[width=16pc,height=14pc]{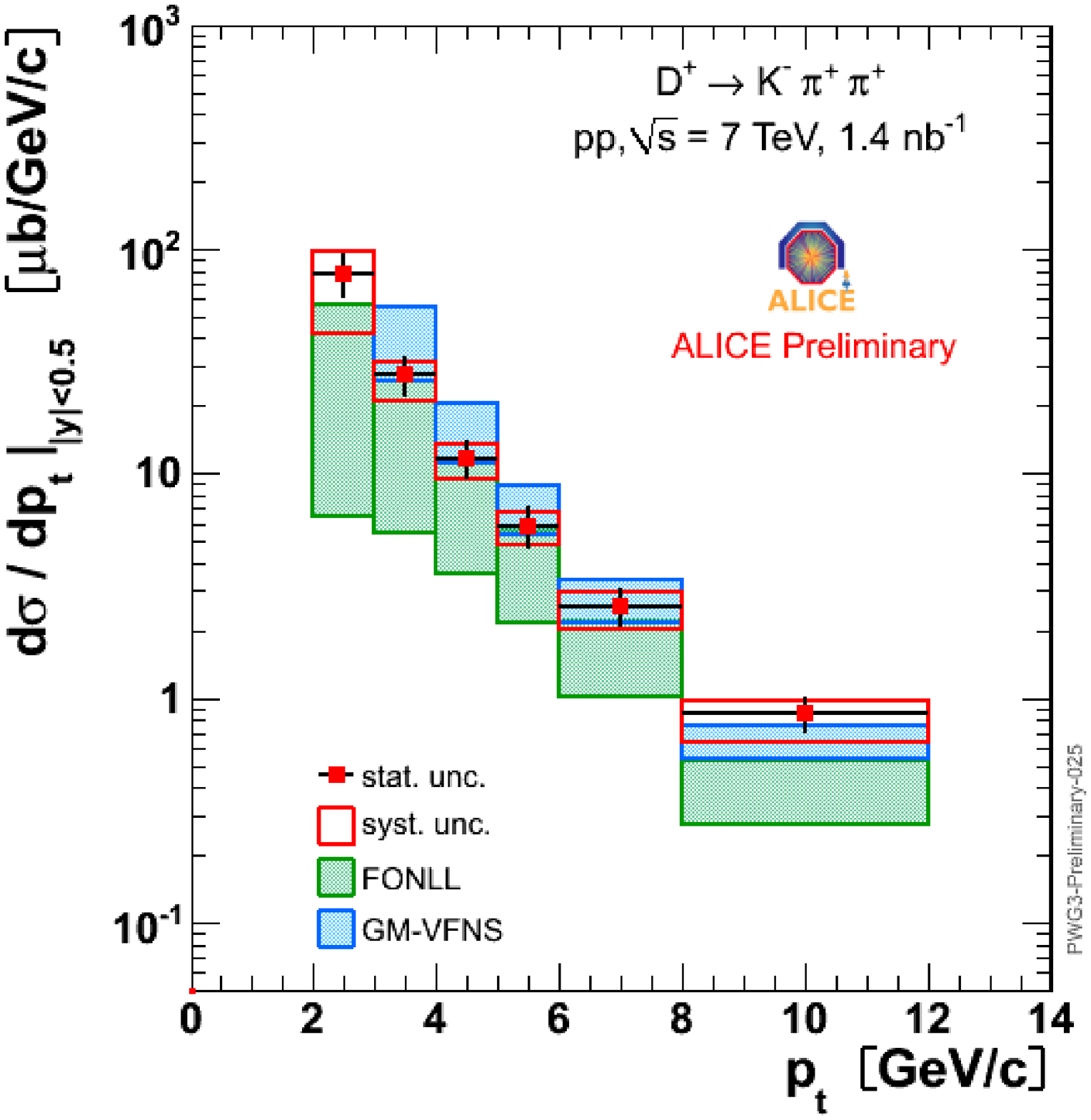}
\caption{ Preliminary $\rm p_{t}$ differential cross section  for $\rm D^{0}$ (left panel) and
$\rm D^{+}$ (right panel) compared with FONLL \cite{cacciari} and
GM-VFNS \cite{GM} theoretical predictions}
\label{sigma}
\end{figure}
Figure \ref{sigma}  shows the first preliminary $ p_t$ differential
cross section for $\rm D^{0}$ and $\rm D^{+}$ mesons,
 compared with  two theoretical predictions, FONLL \cite{cacciari} and
 GM-VFNS \cite{GM}. The
 measurements are well described by both models within their
 theoretical uncertainties.
With the full statistics of 2010, we expect to increase
the $p_{t}$ coverage.

\begin{figure}[ht]
\includegraphics[width=16pc,height=13pc]{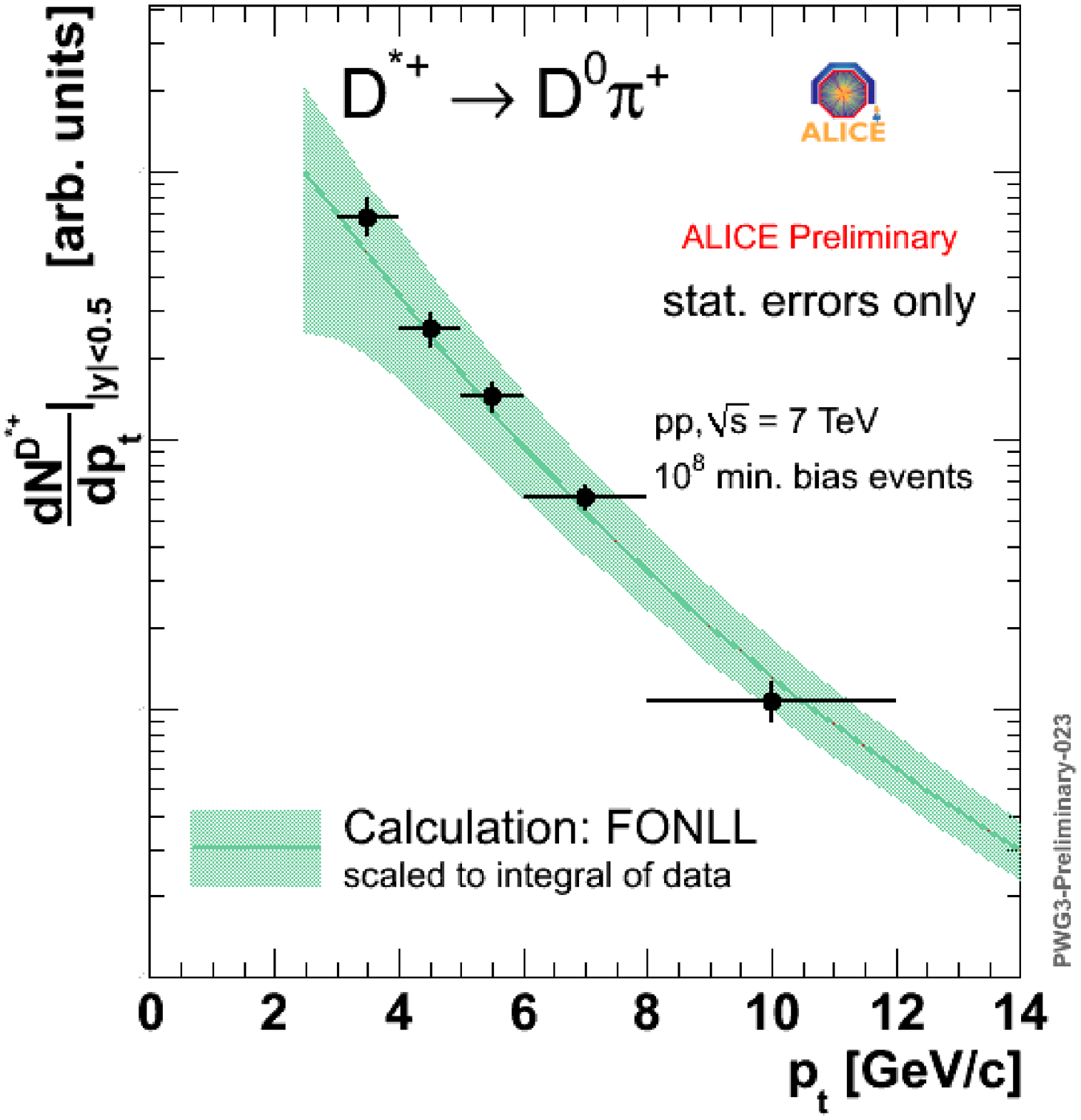}
\includegraphics[width=16pc,height=13pc]{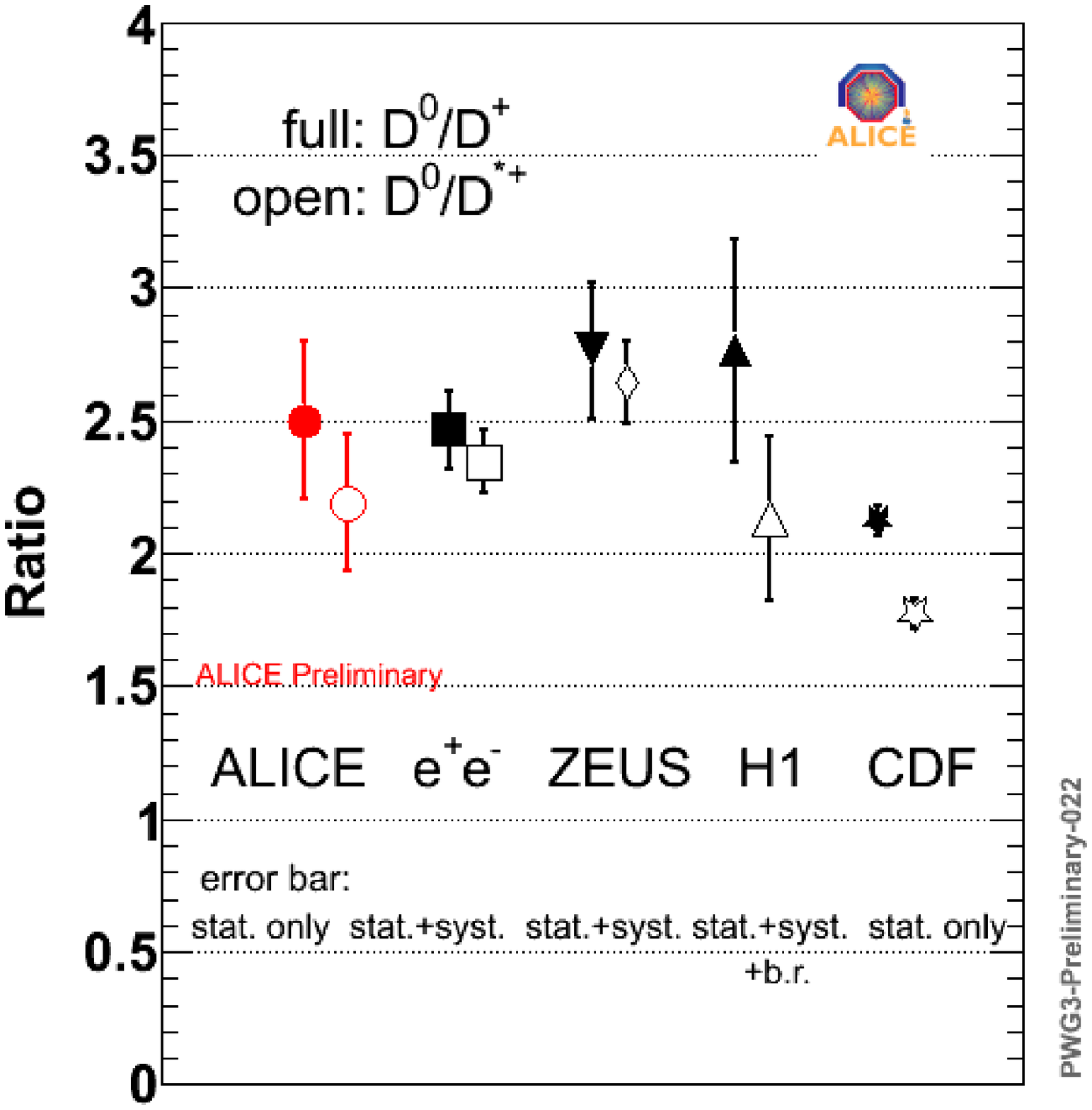}
\caption{Left Panel: $p_t$ distribution of $\rm D^{*+}$ meson, shown with
  statistical errors only. Right Panel: D-mesons ratio for $ p_{t}>$3
  GeV/c compared to previous measurements }
\label{dndpt}
\end{figure}

The $\rm D^{*+} \rightarrow D^{0} \pi^{+}$ analysis is also well
advanced, with the signal extracted in various $p_{t}$
bins. Figure \ref{dndpt} (right panel) shows the $p_{t}$
distribution of $\rm D^{*+}$ mesons in the
range 3$< p_{t} <$ 12 GeV/c in arbitrary units, with statistical errors
only (evaluation of systematic uncertainties is ongoing) and
compared to the shape of FONLL theoretical predictions
\cite{cacciari}.  The $p_{t}$ shape is well described by the pQCD
predictions. For $p_{t}>$3 GeV/c, we have derived the D-meson
ratios ($\rm D^{0}/D^{+}$ and $\rm D^{0}/D^{*+}$) that are found to be in
agreement with previous measurements  at lower
energies \cite{Zeus,CDF} as shown in the right panel of figure  \ref{dndpt}.
Promising signals have also been observed for the other decay channels
$\rm D^{0} \rightarrow K^{-} \pi^{+} \pi^{-} \pi^{+}$, $\rm D^{+}_{s}
\rightarrow K^{-} K^{+} \pi^{+}$ and $\rm \Lambda_{c}^{+} \rightarrow p
K^{-} \pi^{+}$.

\section{Conclusions}The ALICE detector provides excellent tracking,
vertexing and particle identification capabilities that  allow a high precision measurement of the open
charm cross section via hadronic decays. We have shown the preliminary
results on the measurement of  the production cross section of the
$\rm D^{0}$ and $\rm D^{+}$ mesons at central rapidity in pp collisions at
$\sqrt{s}$=7 TeV  in 2 $<p_{t} <$
 10 GeV/c. The measurements are described by the theoretical
 calculations within their  uncertainties. These measurements will
 provide reference data to measure the energy loss of D-mesons in Pb--Pb
 collisions at the  LHC.

\end{document}